\documentclass[useAMS,usenatbib]{mn2e}
\usepackage{natbib}
\usepackage{amsmath,amsfonts}
\usepackage{epsfig}
\usepackage{mathptmx}

\def\reff@jnl#1{{\rm#1\/}}

\def\aj{\reff@jnl{AJ}}                  
\def\araa{\reff@jnl{ARA\&A}}            
\def\apj{\reff@jnl{ApJ}}                        
\def\apjl{\reff@jnl{ApJ}}               
\def\apjs{\reff@jnl{ApJS}}              
\def\ao{\reff@jnl{Appl.Optics}}         
\def\apss{\reff@jnl{Ap\&SS}}            
\def\aap{\reff@jnl{A\&A}}               
\def\aapr{\reff@jnl{A\&A~Rev.}}         
\def\aaps{\reff@jnl{A\&AS}}             
\def\azh{\reff@jnl{AZh}}                        
\def\baas{\reff@jnl{BAAS}}              
\def\jrasc{\reff@jnl{JRASC}}            
\def\memras{\reff@jnl{MmRAS}}           
\def\mnras{\reff@jnl{MNRAS}}            
\def\pra{\reff@jnl{Phys. Rev. A}}         
\def\prb{\reff@jnl{Phys. Rev. B}}         
\def\prc{\reff@jnl{Phys. Rev. C}}         
\def\prd{\reff@jnl{Phys. Rev. D}}         
\def\prl{\reff@jnl{Phys. Rev. Lett}}      
\def\pasp{\reff@jnl{PASP}}              
\def\pasj{\reff@jnl{PASJ}}              
\def\qjras{\reff@jnl{QJRAS}}            
\def\skytel{\reff@jnl{S\&T}}            
\def\solphys{\reff@jnl{Solar~Phys.}}    
\def\sovast{\reff@jnl{Soviet~Ast.}}     
\def\ssr{\reff@jnl{Space~Sci.Rev.}}     
\def\zap{\reff@jnl{ZAp}}                        
\def\nat{\reff@jnl{Nature}}             

\def\p#1by#2{{\partial{#1} \over \partial{#2}}}
\def\pp#1by#2#3{{\partial^2{#1} \over \partial{#2}\partial{#3}}}
\def\d#1by#2{{{\rm d}{#1} \over {\rm d}{#2}}}
\def\dd#1by#2#3{{{\rm d}^2{#1} \over {\rm d}{#2}{\rm d}{#3}}}

\title[Spinning dust emission in NGC~6946]{Microwave observations of spinning dust emission in NGC~6946\thanks{We request that any reference to this paper cites ``AMI Consortium: Scaife et~al. 2010''.}}

\label{firstpage}

\author[Scaife et~al.]{
 AMI Consortium: Anna M. M. Scaife$^1$\thanks{email: ascaife@cp.dias.ie}, 
 Bojan Nikolic$^{2,3}$, 
 David A. Green$^3$,
 Rainer Beck$^4$,  
\newauthor
 Matthew L.\ Davies$^3$, 
 Thomas M.\ O.\ Franzen$^3$,
 Keith J. B. Grainge$^{2,3}$, 
 Michael P.\ Hobson$^3$, 
\newauthor 
 Natasha Hurley-Walker$^3$, 
 Anthony N.\ Lasenby$^{2,3}$, 
 Malak Olamaie$^3$, 
 Guy G. Pooley$^3$, 
\newauthor
 Carmen Rodr{\'i}guez-Gonz{\'a}lvez$^3$, 
 Richard D.\ E.\ Saunders$^{2,3}$, 
 Paul F.\ Scott$^3$,
\newauthor
 Timothy W.\ Shimwell$^3$, 
 David J.\ Titterington$^3$, 
 Elizabeth M.\ Waldram$^3$
 \& Jonathan T.\ L.\ Zwart$^5$\\
$^1$ Dublin Institute for Advanced Studies, 31 Fitzwilliam Place,
     Dublin 2, Ireland\\
$^2$ Kavli Institute for Cosmology Cambridge, Madingley Road,
     Cambridge, CB3 0HA\\
$^3$ Astrophysics Group, Cavendish Laboratory, J J Thomson Avenue,
     Cambridge CB3 0HE\\
$^4$ MPIfR, Auf dem H{\"u}gel 69, 53121 Bonn, Germany\\
$^5$ Columbia Astrophysics Laboratory, Columbia University, 550 West 120th
     Street, New York 10027, USA}

\date{Accepted ---; received ---; in original form \today}

\pagerange{\pageref{firstpage}--\pageref{lastpage}}

\pubyear{2009}

\begin{document}
\maketitle

\begin{abstract}
We report new cm-wave measurements at five frequencies between 15 and 18\,GHz of the continuum emission from the reportedly anomalous ``region 4'' of the nearby galaxy NGC~6946. We find that the emission in this frequency range is significantly in excess of that measured at 8.5\,GHz, but has a spectrum from 15--18\,GHz consistent with optically thin free--free emission from a compact {\sc Hii} region. In combination with previously published data we fit four emission models containing different continuum components using the Bayesian spectrum analysis package {\tt radiospec}. These fits show that, in combination with data at other frequencies, a model with a spinning dust component is slightly preferred to those that possess better-established emission mechanisms.

\end{abstract}

\begin{keywords}
radiation mechanisms: general --- galaxies: individual(NGC~6946)
\end{keywords}

\section{Introduction}

The complete characterization of microwave emission from spinning dust
grains is a key question in both astrophysics and cosmology as it probes a
region of the electromagnetic spectrum where a number of different
astrophysical disciplines overlap: it is important for CMB
observations in order to correctly characterise the contaminating foreground
emission (Gold et~al. 2010); for star and planetary formation it is important because it
potentially probes a regime of grain sizes that is not otherwise easily
observable (Rafikov 2000).

Although a number of objects have now been found to exhibit anomalous
microwave emission, attributed to spinning dust, it is still unclear
what differentiates those objects from the many other seemingly
similar targets that do not show the excess. In order to investigate this question a number of Galactic observations have been made towards known star formation regions (see e.g. AMI Consortium: Scaife et~al. 2010 and references therein; Casassus et~al. 2008; Tibbs et~al. 2009). In addition Murphy et~al. (2010; hereinafter M10) made the first extra-galactic search for anomalous microwave emission within the star formation regions of the nearby galaxy NGC~6946 using the Caltech Continuum Back-end on the Green Bank Telescope. M10 found a significantly anomalous spectrum in only one of 10 star-forming regions: extra-nuclear region 4 (hereinafter NGC~6946-E4). The excess of emission was seen between 27--38\,GHz relative to the continuum emission at 8.5\,GHz measured using combined Effelsberg 100\,m Telescope and VLA observations (Beck 2007).

In this letter we present follow-up observations of NGC~6946-E4 at frequencies from 15-18\,GHz using the Arcminute Microkelvin Imager (AMI) Large Array (LA). In Section~\ref{sec:obs} we present the details of these observations, in Section~\ref{sec:results} we present the results of the AMI-LA observations and a comparison with other radio data, and in Section~\ref{sec:conc} we discuss the implications of these results and form our conclusions. In what follows we use the convention: $S\propto \nu^{\alpha}$, where $S$ is flux density, $\nu$ is frequency and $\alpha$ is the spectral index. All errors are quoted to 1\,$\sigma$. 

\section{Observations}
\label{sec:obs}

The AMI-LA consists of eight 13\,m antennas and is 
 sited at Lord's Bridge, Cambridge (AMI Consortium: Zwart
et~al. 2008). The telescope observes in 
the band 13.9--18.2\,GHz with cryogenically cooled NRAO indium-phosphide
front-end amplifiers. The overall system temperature is approximately
25\,K. Amplification, equalization, path compensation and
automatic gain control are applied to the IF signal. The
back-end has an analogue lag correlator with 16 independent 
correlations formed at double the Nyquist rate for each baseline using path delays spaced by 26\,mm. From these, eight complex visibilities are formed in 
0.75\,GHz bandwidth channels. In what follows, the three
lowest frequency channels (1, 2 \& 3; 13.9--14.6\,GHz) are not used due to interference from geostationary satellites. 

NGC~6946-E4 (J\,$20^{\rm{h}} 34^{\rm{m}} 19\fs 17\,+60\degr 10\arcmin 08\farcs7$) was observed by the AMI-LA in a single 12 hour synthesis. Data reduction was performed using the local software tool \textsc{reduce}. This applies
both automatic and manual flags for interference, 
shadowing and hardware errors, phase and amplitude
calibrations and then
Fourier transforms the correlator data to synthesize frequency
channels 
before output to disk in \emph{uv} FITS format suitable for imaging in
\textsc{aips}. Flux calibration was performed using short observations of 3C48 at the beginning of the run and 3C286 at the end of the run. We assumed I+Q flux densities for these sources in the
AMI-LA channels consistent with the frequency dependent model of Baars et al. (1977), $\simeq$ 1.64 and 3.48\,Jy, respectively, at
15\,GHz. As Baars et~al. measure I and AMI-LA measures
I+Q, these flux densities 
include corrections for the polarisation of the calibrator sources derived
by interpolating from VLA 5, 8 and 22\,GHz observations. A correction is
also made for the changing intervening airmass over the observation. From
cross-calibration of 3C48 and 3C286, we find the flux calibration is accurate to better than
5 per cent. The phase was calibrated using interleaved observations of J\,2031+5455,  
selected from the Jodrell Bank VLA Survey (JVAS; Patnaik et~al. 1992). After calibration, the phase is generally stable to
$5^{\circ}$ for channels 4--7, and
$10^{\circ}$ for channel 8. The FWHM of the primary beam of the AMI-LA is $\approx 6$\arcmin at
16\,GHz. 

Reduced data were imaged using the {\sc aips} data package. {\sc{clean}}
deconvolution was performed using the task 
{\sc{imagr}}. {\sc{clean}} deconvolution maps were made from both the
combined channel set and for individual channels.  

\section{Results}
\label{sec:results}

\begin{figure*}
\includegraphics[clip=,width=0.4\textwidth]{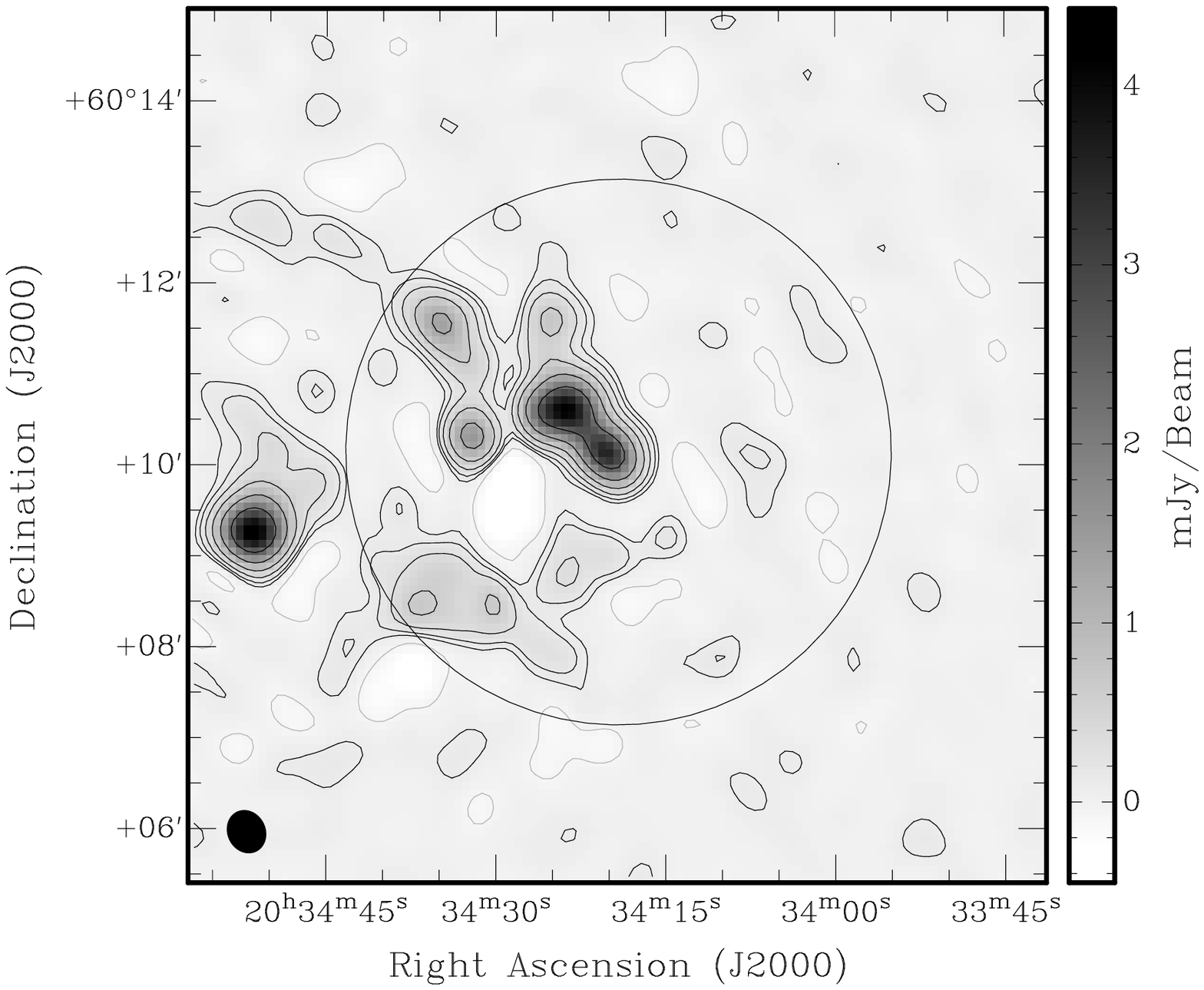}\qquad \includegraphics[clip=,width=0.405\textwidth]{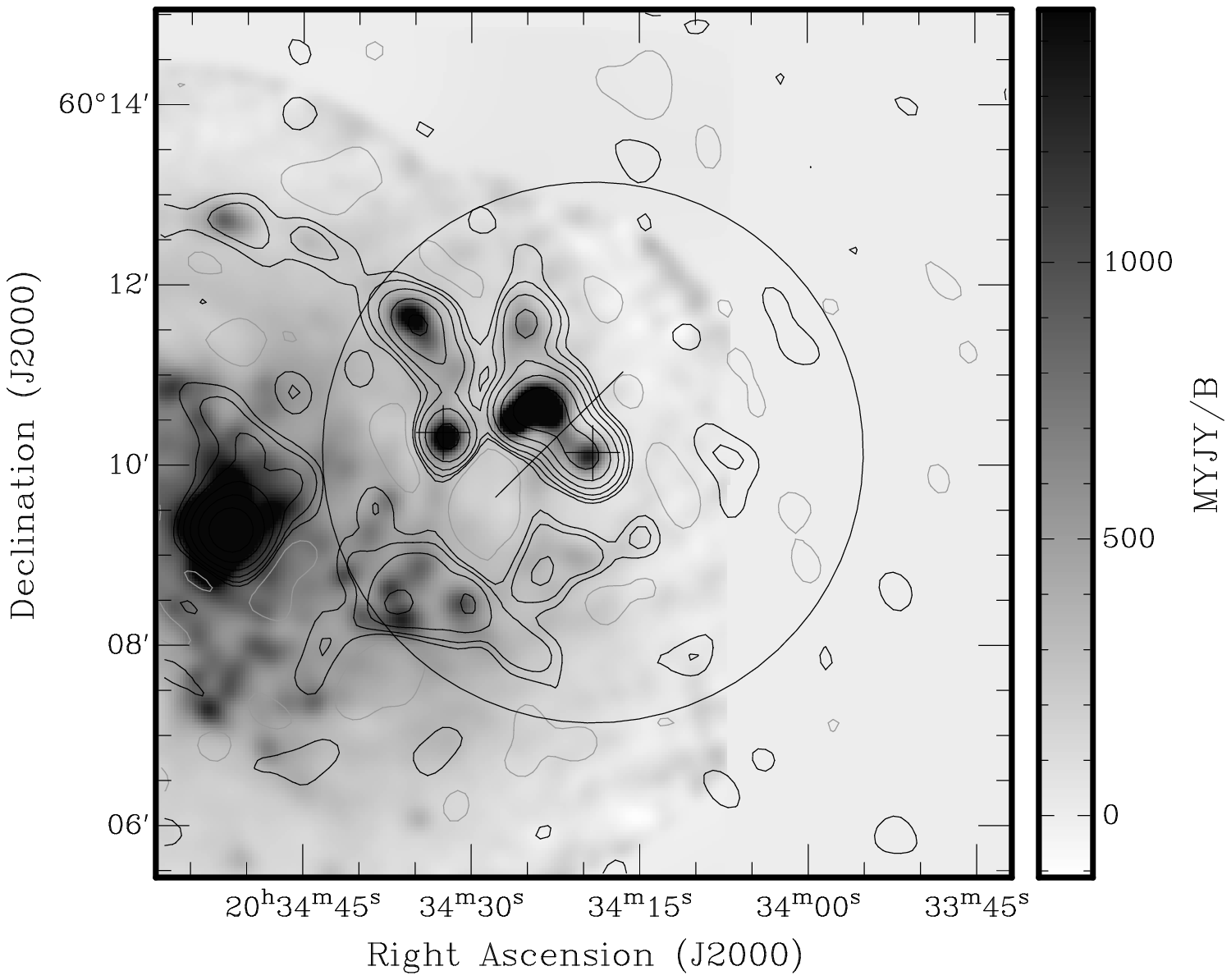}
\centerline{(a)\hspace{0.4\textwidth}(b)}
\caption{NGC~6946-E4. (a) Greyscale and contours at $-3,3,6,12\,\sigma$ etc are from the AMI-LA where $\sigma_{\rm{rms}} = 25\,\mu$Jy\,beam$^{-1}$, greyscale units are mJy\,beam$^{-1}$. Data are not corrected for the primary beam response. The AMI-LA synthesized beam, $29.2''\times25.2''$, is shown as a filled ellipse in the bottom left corner. (b) Contours are from the AMI-LA and greyscale at 8.5\,GHz is from Effelsberg-VLA measurements, greyscale units are $\mu$Jy\,beam$^{-1}$ (Beck 2007). The positions of NGC~6946-E4 (at the phase centre) and NGC~6946-E8 (to the east) are marked with crosses, and a line is drawn to indicate a non-fitted polygon edge (see text for details). The AMI-LA primary beam is shown as a black circle in both images. \label{fig:amimap}}
\end{figure*}

\subsection{AMI Large Array Data}

The combined channel map from the AMI-LA observation of NGC~6946-E4 is shown in Fig.~\ref{fig:amimap}. E4 is located at the phase centre with extra-nuclear region 8 (J\,$20^{\rm{h}} 34^{\rm{m}} 32\fs 52\,+60\degr 10\arcmin 22\farcs 0$; NGC~6946-E8) to the east and  the galactic nucleus of NGC~6946 (J\,$20^{\rm{h}} 34^{\rm{m}} 52\fs 34\,+60\degr 09\arcmin 14\farcs 2$) further to the east, outside the primary beam FWHM. Flux densities were extracted for both NGC~6946-E4 and NGC~6946-E8. Since NGC~6946-E8 also lies within the FWHM of the primary beam we used the recovered flux densities for this source as a check on the absolute calibration for this field. The flux densities measured for NGC~6946-E4 and NGC~6946-E8 are listed in Table~\ref{tab:fluxes}.

\subsection{NCC6946-E4: Comparison with other radio data}
\begin{table*}
\begin{center}
\caption{Flux densities.\label{tab:fluxes}}
\begin{tabular}{lcccccc}
\hline\hline
&& \multicolumn{5}{c}{AMI-LA Channel Number}\\
\cline{3-7}
Source & 8.5\,GHz & 4 & 5 & 6 & 7 & 8 \\
       & (mJy)& (mJy) & (mJy)& (mJy)& (mJy)& (mJy)\\
    & [2] & [3] & [4] & [5] & [6] & [7]  \\
\hline
NGC~6946-E4 & $1.94\pm0.29$ & $3.17\pm0.36$ & $3.16\pm0.33$ & $3.00\pm0.28$ & $3.15\pm0.43$ & $2.78\pm0.34$ \\
           &  & &  & & ($3.31\pm0.43$) & ($3.09\pm0.35$) \\
NGC~6946-E8 & $2.11\pm0.21$ & $1.90\pm0.27$ & $1.51\pm0.27$ & $1.40\pm0.23$ & $1.51\pm0.34$ & $1.42\pm0.19$\\
\hline
\end{tabular}
\vskip .05in
\begin{minipage}{0.9\textwidth} 
{\footnotesize Column [1] contains the source name, [2] the \emph{uv}-sampled 8.5\,GHz flux density for that source, [3]-[7] contain the AMI-LA flux densities for each source from AMI-LA channels 4--8, with flux loss corrected values in parentheses..
}
\end{minipage}
\end{center}
\end{table*}

We compared the AMI-LA data with that at 8.5\,GHz from the Effelsberg 100\,m and VLA telescopes (Beck 2007). These data in their original form, see Fig.~\ref{fig:amimap}(b), constitutes a total power measurement of the region at a resolution of $15''$. We Fourier transformed these combined data and sampled them at \emph{uv} positions identical to those of the AMI-LA 15.0\,GHz data (channel 4) in order to compare consistent angular scales. The resulting \emph{uv} data set was then mapped and cleaned in the same way as the AMI-LA data. Since channel 4 of the AMI-LA recovers a large amount of the extended structure around NGC~6946-E4 the flux density for this source from the sampled 8.5\,GHz map is very similar to the unsampled value, see Table~\ref{tab:fluxes}. 
Flux densities were extracted from both the AMI-LA and sampled Effelsberg 8.5\,GHz maps using the {\sc fitflux} software (Green 2007; AMI Consortium: Scaife et~al. 2009). This method calculates flux densities by removing a tilted plane fitted to the local background and integrating the remaining flux. We do this by drawing a polygon around the source and fitting a tilted plane to the pixels around the edge of the polygon. Where an edge of the polygon crosses a region confused by another source the background is subjective and we omit this edge from the fitting. An example of where this might be appropriate is shown in Fig.~\ref{fig:amimap} (b). Since the extracted flux density is dependent to some degree on the background emission we repeat this process using five irregular polygons, varying each slightly in shape. The final flux density is the average of that extracted from these five polygons. 

Errors on the flux densities were calculated as $\sigma_{\nu} = \sqrt{(0.05S_{\nu})^2+\sigma_{\rm{rms},\nu}^2+\sigma_{\rm{fit}}^2}$, where $\sigma_{\rm{rms},\nu}$ is the rms noise outside the primary beam on each channel map and $\sigma_{\rm{fit}}$ is the standard deviation of the fluxes measured in the five polygonal apertures. The errors are dominated by $\sigma_{\rm{fit}}$, which is large due to the complicated background emission in this crowded field.

Since the \emph{uv} coverage of the AMI-LA varies across the frequency channels we quantified the amount of flux lost in each channel relative to channel 4 by sampling the total power 8.5\,GHz map to match the \emph{uv} coverage in each of the AMI-LA frequency channels. This showed that flux loss was negligible in channels 4--6 and notable ($\ge5\%$) in channels 7 \& 8 only. We corrected for this loss in the measured AMI-LA flux densities with corrected values shown in parentheses in Table~\ref{tab:fluxes}. In the same manner we checked for corrections to the flux densities of NGC~6946-E8. In this case the corrections were negligible ($<2\%$) and no corrections were made. We found the 15--18\,GHz data for NGC~6946-E8 to be consistent with the spectrum for this object presented in M10. The flux density of NGC~6946-E4 across the AMI-LA band is in excess of the 8.5\,GHz flux density by approximately 10\,$\sigma$. 

The collected observations of both region E4 and E8 were analysed
using an updated version of the {\tt radiospec} package\footnote{The complete code and data used for this spectrum analysis are
available for public download under GPL license: 
{\tt http://www.mrao.cam.ac.uk/$\sim$bn204/galevol/speca/sdgals.html}.
}
\citep{2009arXiv0912.2317N}. This software tool calculates the
Bayesian posterior distribution of the parameters of a model for the
radio spectrum and, furthermore, the Bayesian evidence. The implementation of
these calculations is based on the nested sampling algorithm by
Skilling (2004). For the analysis of the data in this paper, we
used several different models which each consist of a number of
components, each with physically parameterized properties. Two components that are present in all models are a synchrotron
component, which we parametrise in terms of the supernova rate within
the beam, and an un-absorbed free--free component, which we
parametrise in terms of the star-formation rate within the beam. The
conversions from these physical parameters to radio luminosities are
made according to the formulae given by Condon (1992).

In order to explain the excess of emission at cm wavelengths, it is
necessary to introduce another component that contributes to the
emission. As outlined above we have considered two options: emission
due to spinning dust, and absorbed free--free emission from a compact
H\,{\sc ii} region. For the spinning dust emission component we
 used the warm ionized medium model described by \cite{1998ApJ...508..157D}\footnote{
 { \tt ftp://ftp.astro.princeton.edu/draine/dust/\\spin/emit4.jnu.wim\_a}}. The only degree of
freedom in this model is the overall amplitude, which we parametrised
in terms of the total mass of gas carrying the spinning dust.

Our model for the absorbed free--free emission from H\,{\sc ii} is
again parametrised in terms of the star-formation rate in the region,
but it now also has the filling factor, i.e. the fraction of the area covered by the telescope beam that the H\,{\sc ii} region subtends, as a free parameter. This area is used to compute the electron opacity as a function of frequency and to correct the unabsorbed free--free model for the effects of absorption.

\begin{figure*}
\begin{tabular}{cc}
Synchrotron + free--free + absorbed free--free &
Synchrotron + free--free + spinning dust\\
\includegraphics[width=0.4\textwidth]{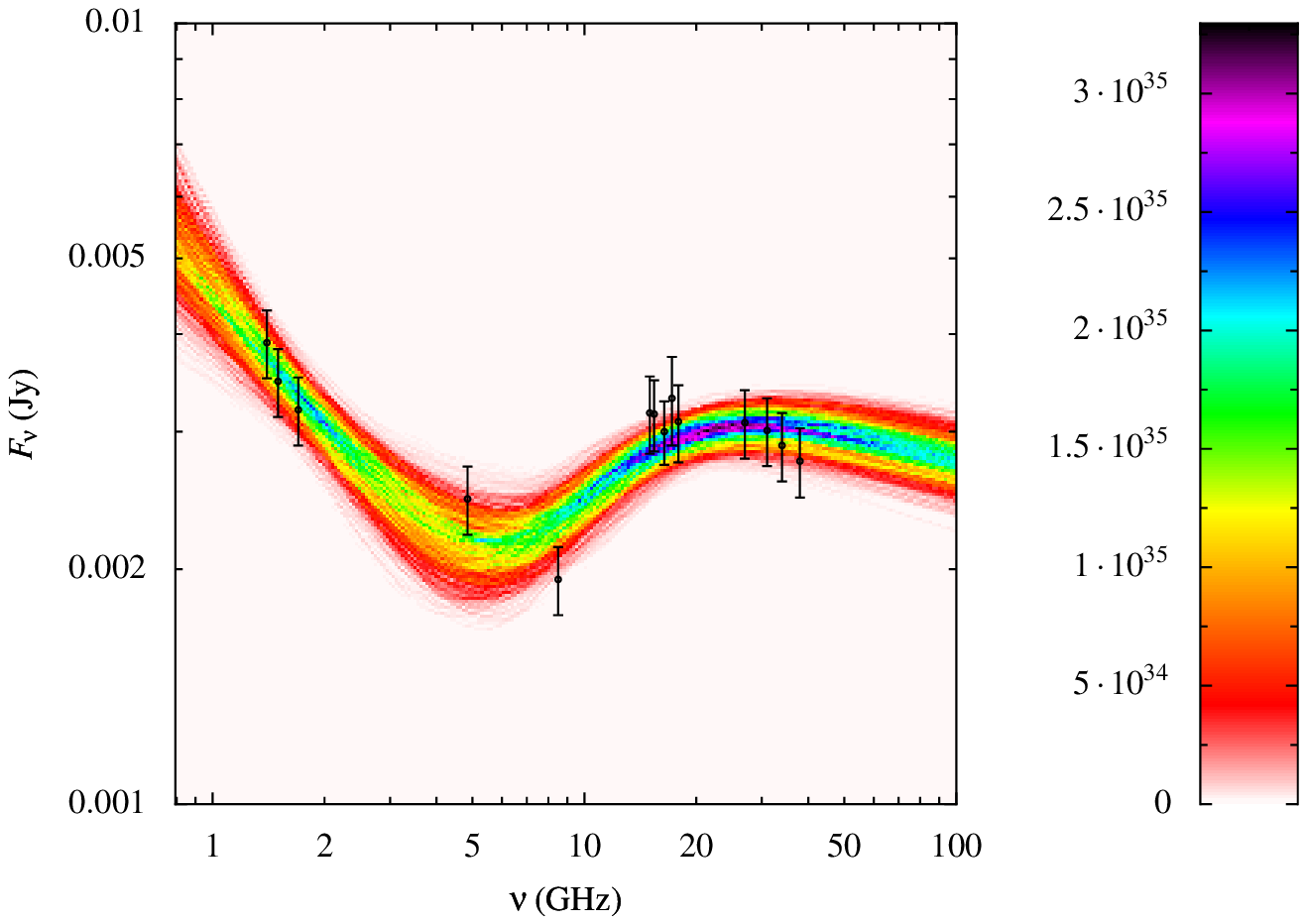}
&
\includegraphics[width=0.4\textwidth]{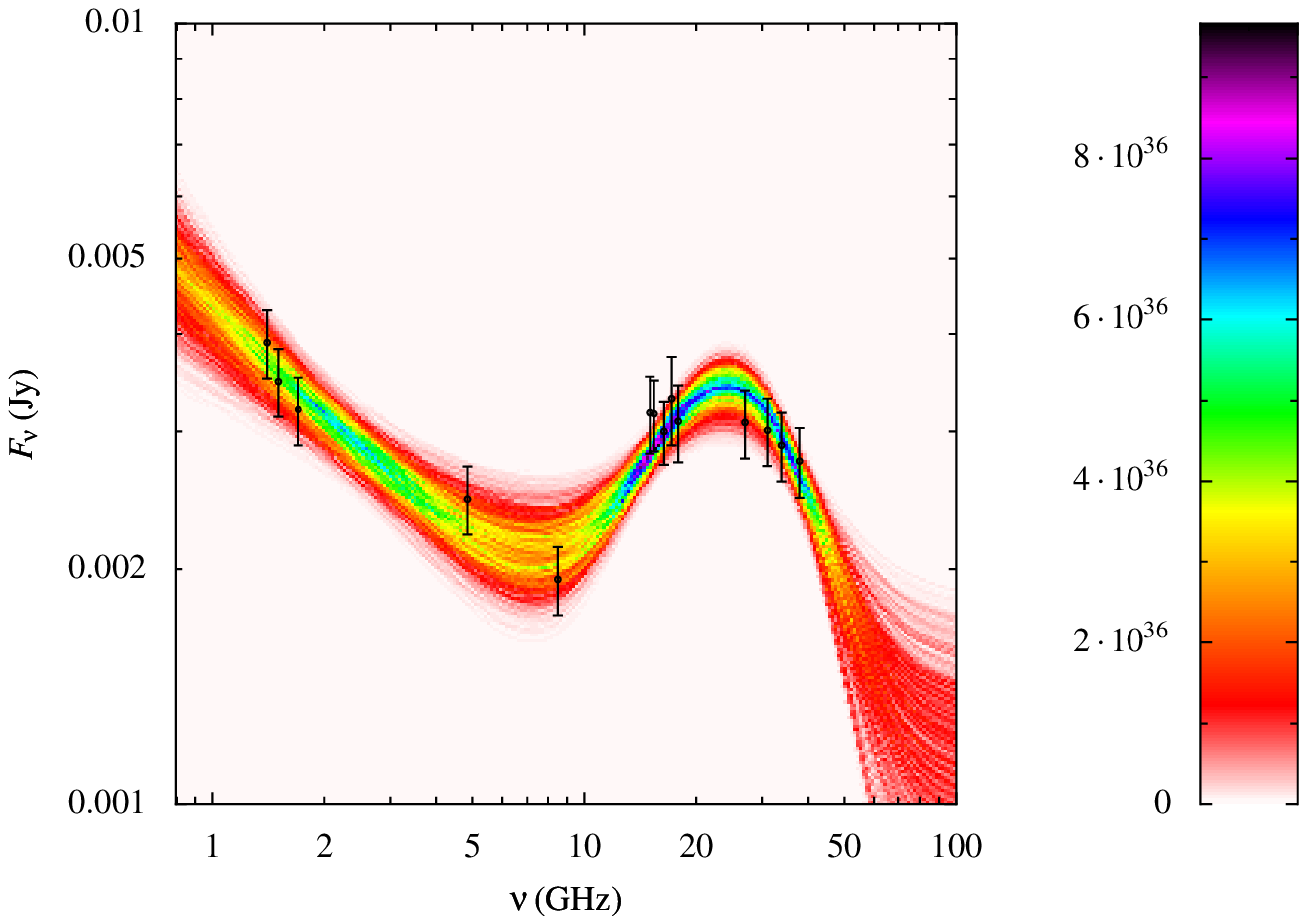}
\end{tabular}
\caption{The observed radio spectrum of region E4 of NGC~6946 (points
  and error bars) with the fan-diagram of two model fits to these
  data: on the left is the model with a highly absorbed free--free
  emission region (H1) and on the right is the model with spinning dust
  emission (H2).  Low frequency data are taken from M10, scaled to the
  \emph{uv}-sampled flux density at 8.5\,GHz, with the exception of
  points between  15 and 18\,GHz which are from the AMI-LA (this work). The
  assumed error is 10\% unless stated otherwise in
  Table~\ref{tab:fluxes}. The colour scale indicates the evidence contribution as a function of frequency and flux density, for details see Nikolic 2009.}
\label{fig:E4spec}
\end{figure*}

\begin{figure*}
\begin{tabular}{cc}
Synchrotron + free--free  &
Synchrotron + free--free + spinning dust\\
\includegraphics[width=0.4\textwidth]{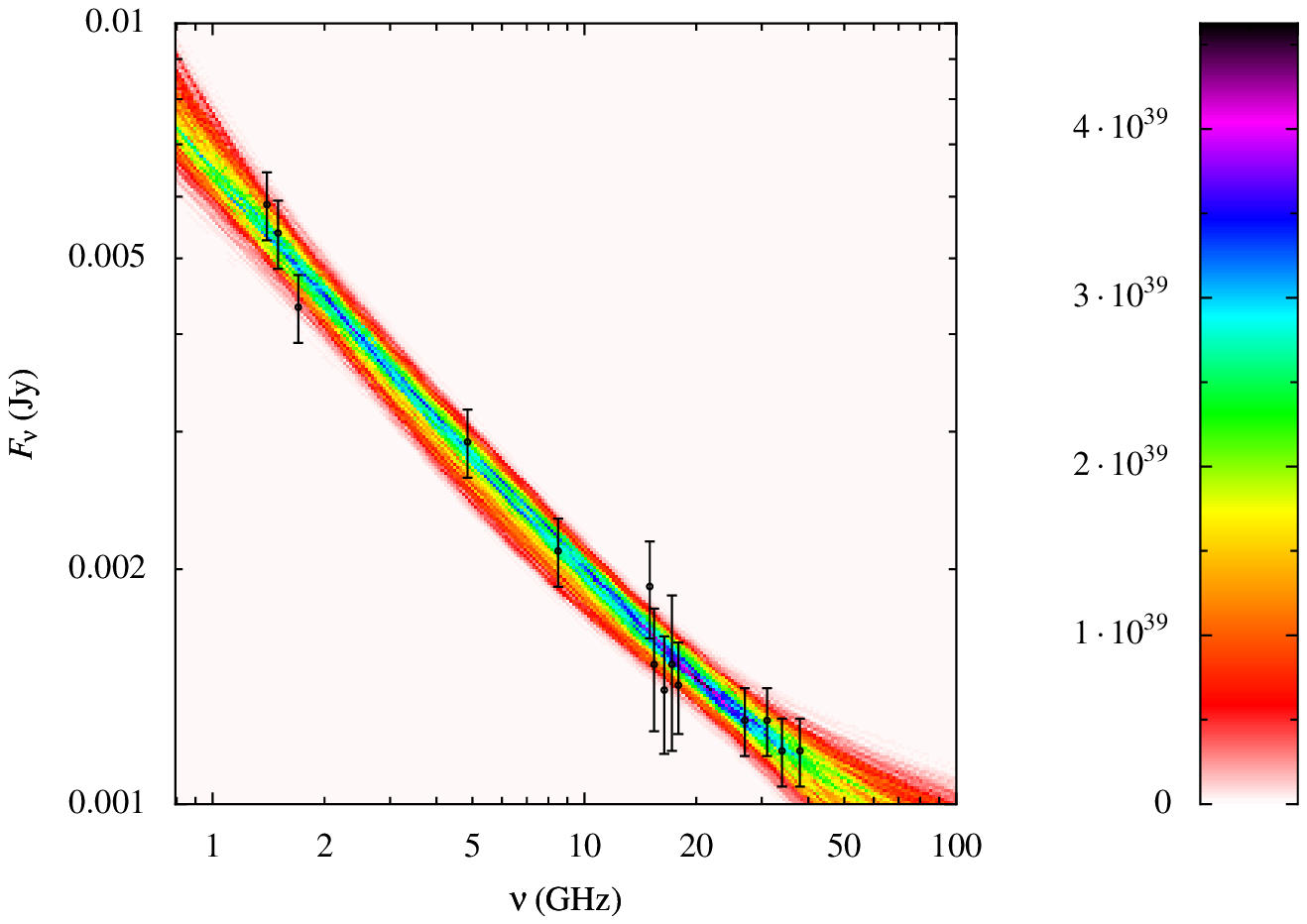}&
\includegraphics[width=0.4\textwidth]{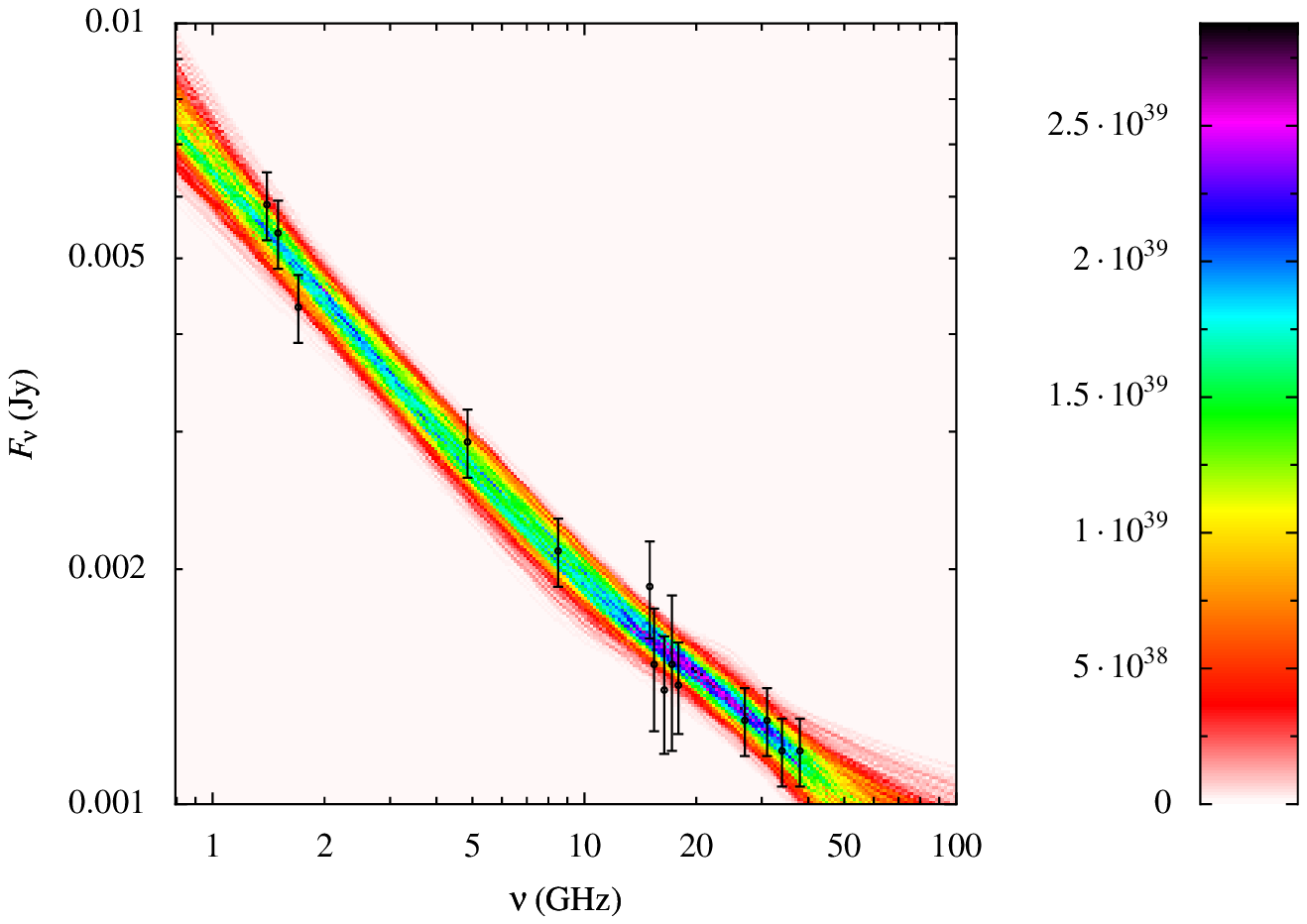}
\end{tabular}
\caption{The observed radio spectrum and fan-diagrams for region E8 of
  NGC~6946, with data and errors as in Fig.~\ref{fig:E4spec}. The
  model on the left only consists of a synchrotron and un-absorbed free--free
  components (H1) while the model on the right also has a spinning dust
  component (H2). Low frequency data are taken from M10, scaled to the
  \emph{uv}-sampled flux density at 8.5\,GHz, with the exception of
  points between  15 and 18\,GHz which are from the AMI-LA (this work). The
  assumed error is 10\% unless stated otherwise in Table~\ref{tab:fluxes}. Colour scale as above.}
\label{fig:e8specfan}
\end{figure*}

\begin{figure*}
\begin{tabular}{cc}
Region E4&
Region E8\\
\includegraphics[width=0.4\textwidth]{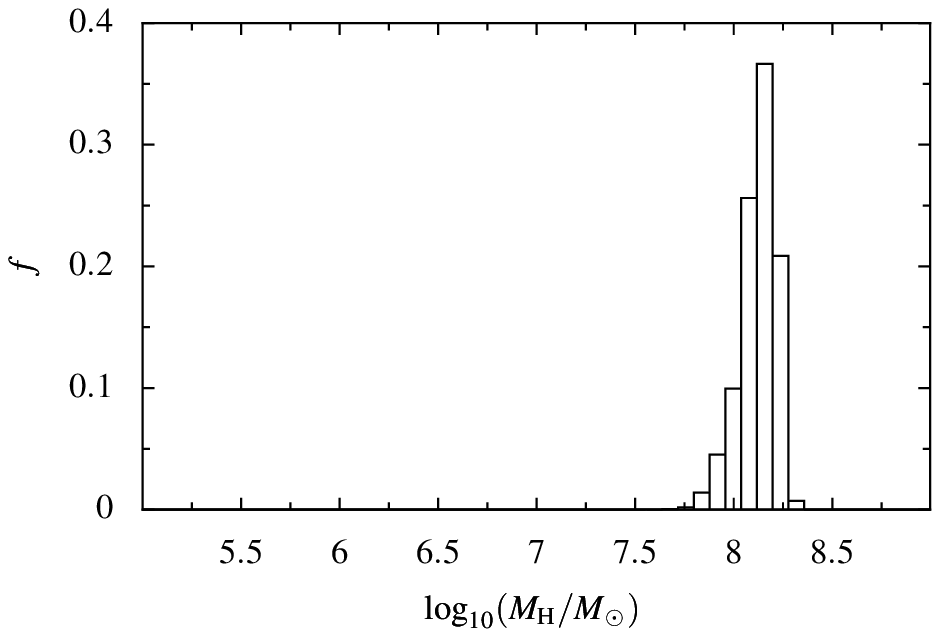}&
\includegraphics[width=0.4\textwidth]{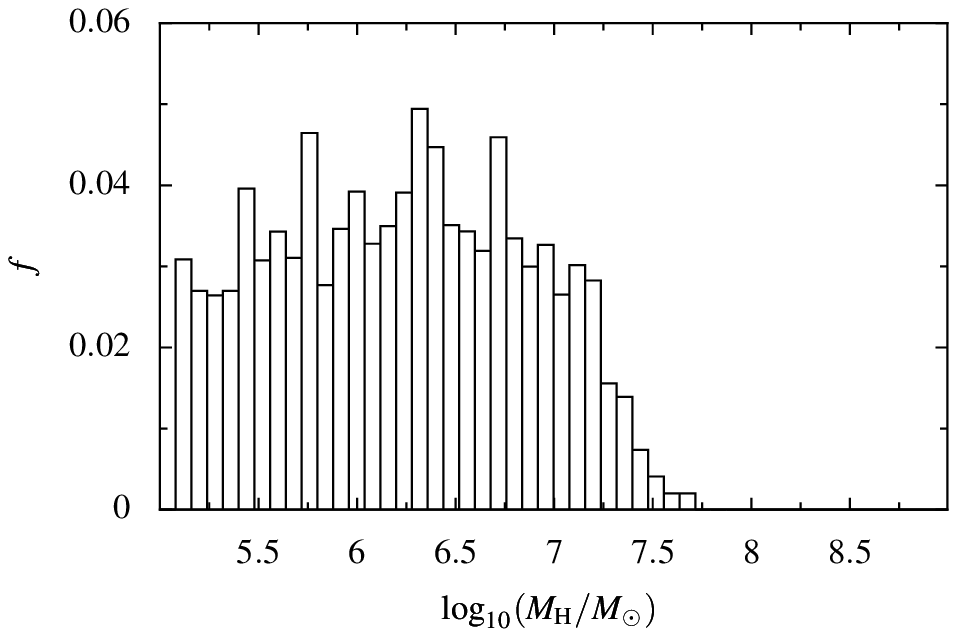}
\end{tabular}
\caption{Marginalised distribution of inferred gas mass that bears the
spinning dust: the left panel shows extra-nuclear region 4 and the right panel extra-nuclear region 8.}
\label{fig:sdmass-margin}
\end{figure*}


\begin{table*}
\begin{center}
\caption{Derived model parameters and errors \label{tab:parms}}
\begin{tabular}{lcccccc}
\hline\hline
Model & SNe rate & $\alpha_{\rm{sync}}$ & $SFR_{\rm{abs}}$ & $SFR_{\rm{unabs}}$ & $M_{\rm{gas}}$ & $f$ \\
      & $\log_{10}(\rm{yr}^{-1})$ & & $\log_{10}(\rm{yr}^{-1})$ & $\log_{10}(\rm{yr}^{-1})$ & $\log_{10}(\rm{M_{\odot}})$ & $\log_{10}(\rm{str})$ \\
      & [2] & [3] & [4] & [5] & [6] & [7] \\
\hline 
\emph{Prior} & $ (-5,-1)$ & $(-1.0,-0.5)$ & $(-3,-1) $ & $(-3,0)$ & $(5,9)$ & $(-6,-3)$ \\
NGC~6946-E4 (H1) & $-3.84\pm0.11$ & $-0.74\pm0.14$ & $-0.93\pm0.07$ & $-1.80\pm0.48$ & -- & $-4.97\pm0.18$ \\
NGC~6946-E4 (H2) & $-3.98\pm0.16$ & $-0.71\pm0.15$ & -- & $-1.40\pm0.36$ &  $8.12\pm0.09$ & -- \\
NGC~6946-E8 (H1) & $-3.65\pm0.04$ & $-0.66\pm0.13$ & -- & $-1.70\pm0.41$ &-- & -- \\
NGC~6946-E8 (H2) & $-3.64\pm0.04$ & $-0.65\pm0.12$ & -- & $-1.75\pm0.41$ & $6.19\pm0.66$ & -- \\  
\hline
\end{tabular}
\end{center}
\end{table*}

\section{Discussion and Conclusions}
\label{sec:conc}
Considered on their own, the AMI-LA data (after correction for flux
loss) have a spectral index of
$\alpha_{\rm{AMI}}=-0.11\pm0.77$. Although this value is consistent with optically thin free--free emission, the error on this estimate
is large and we cannot rule out other mechanisms. Since the spectral index between the Effelsberg-VLA
measurement at 8.5\,GHz and the AMI band is rising ($\alpha_{8.5}^{16}=0.67\pm0.08$, see Fig.~\ref{fig:E4spec}), we need to consider the possibility that region E4 contains one or more compact
{\sc Hii} regions with their opacity reaching unity at approximately
12\,GHz. Such an opacity would require an emission measure of $\simeq
5\times10^8$\,pc\,cm$^{-6}$, assuming $T_{\rm{e}} = 10^4$\,K, and would be appropriate for a compact {\sc Hii} region.

We therefore examine two alternative hypotheses for the emission from
region E4. The first is that the emission is due to the usual diffuse
synchrotron and free--free mechanisms associated with star-formation,
with an additional high-opacity free--free component (Hypothesis 1; H1). The second
hypothesis is that there is a spinning dust component rather than 
high-opacity free--free (H2). A summary of how well these two hypotheses fit
the observed data is shown in the form of fan-diagrams in
Fig.~\ref{fig:E4spec}. As can be seen from this figure, neither of the
hypotheses can be ruled out, although the spinning-dust
appears to somewhat better match the data. This is also confirmed by a simple comparison of the models: assuming flat priors and no a priori difference between the models, the logarithmic Bayes factor is $3.7\pm0.3$ in
favour of the spinning dust model. From the Jeffreys' scale of
evidence (Jeffreys 1961; Kass \& Raftery 1995; see e.g. Efron \& Gous 2001 for further discussion of this scale) this would indicate a weak positive preference for the spinning dust model above the free--free model. The maximum likelihood parameters for these models are listed in Table~\ref{tab:parms}.

For comparison we have also carried out a similar analysis on the
region E8, shown in Fig~\ref{fig:e8specfan}. In this case the two
hypotheses are a simple diffuse synchrotron plus free--free model (H1)
versus the same model with an additional spinning dust component (H2). In
this case the logarithm of the Bayesian evidence ratio is $0.5\pm0.3$ in
favour of the simpler model without the spinning dust. A ratio of this size indicates no perceptible difference between the two models.

In region E4 where a spinning dust model is the preferred
hypothesis, we can marginalise the posterior distribution of the
model parameters to obtain an estimate of the gas mass 
containing the spinning dust, shown as a histogram in the left panel of
Fig.~\ref{fig:sdmass-margin} and in numerical form in
Table~\ref{tab:parms}, i.e., $10^{8.1\pm0.1}\,{\rm M}_\odot$. In
region E8 spinning dust is not the preferred hypothesis but proceeding
with this hypothesis anyway, we find an upper limit for the mass of
the gas bearing the spinning dust, which is around $10^{7.5}\,{\rm
  M}_\odot$. We draw a conclusion that if the conditions in E4 and
E8 are similar, then the mass of any gas bearing spinning dust in E8
must be at least a factor of five smaller than in E4.

Region E4 is located on the dense rim of a ``remarkable'' {\sc Hi} hole (Boomsma et~al. 2008) within NGC~6946. Such an association may be relevent to the differentiation of this star formation region from the eight others found to exhibit no anomalous emission by M10. The hole is remarkable for a number of reasons, notably the almost unbroken symmetry of its dense {\sc Hi} rim, unusual in so large a hole, and the small scale high velocity gas complexes observed in connection with it. 

As mentioned above, the spinning dust model is preferred by the
evidence calculation for NGC~6946-E4, but not at a very high level. Definitive
confirmation of the nature of the emission requires measurements at
frequencies above 50\,GHz where the spinning dust and compact H\,{\sc
  ii} region models have significantly different behaviour. For
example, Fig.~\ref{fig:E4spec} shows that at 100\,GHz the difference
between these two models should be at least a factor of two in
brightness.

In the sub-mm there are data available from SCUBA at 850\,$\mu$m for this region by \citealt{2008ApJS..175..277D}, which might be used to constrain the mass of NGC~6946-E4 and place constraints on the frequency at which the optical depth reaches unity. If a compact {\sc Hii} region is present, with $\tau=1$ at $\nu>8.5$\,GHz it should have a correspondingly high dust mass. From analysis of the SCUBA data we obtained a flux estimate
for region E4 of $S_{850} = 11\pm14$\,mJy\,beam$^{-1}$. However, the errors on
this estimate are too high to allow any useful constraint on
the properties of the thermal dust emission or to calculate a reliable dust mass estimate.

From the existing data the possibility of a spinning dust component in this region cannot be ruled out, but the evidence is not yet definitive. Further observations of this object at frequencies covering the higher frequency minimum between spinning dust emission and thermal dust emission ($\approx 90$\,GHz) would be most useful.

\section{ACKNOWLEDGEMENTS}
We thank the staff of the Lord's Bridge Observatory for their
invaluable assistance in the commissioning and operation of the
Arcminute Microkelvin Imager. The AMI-LA is supported by Cambridge
University and the STFC. AS thanks Thijs van der Hulst for drawing attention to the {\sc Hi} complexes in NGC~6946. CRG, TS, TF, MO and MLD   
acknowledge the support of PPARC/STFC studentships.

\end{document}